\documentstyle[prb,aps,twocolumn]{revtex}

\begin{document}
\draft
\flushbottom
\twocolumn[\hsize\textwidth\columnwidth\hsize\csname @twocolumnfalse\endcsname

\title{Surface energy and stability of stress-driven discommensurate
surface structures}
\author{Emilio Artacho$^1$ and J\"org Zegenhagen$^2$}
\address{
   $^1$Instituto de Ciencia de Materiales Nicol\'as Cabrera and
   Departamento de F\'{\i}sica de la Materia Condensada, C-III
\\
   Universidad Aut\'onoma de Madrid, 28049 Madrid, Spain.
\\
   $^2$Max-Planck-Institut f\"ur Festk\"orperforschung,
   Heisenbergstr. 1, 70569 Stuttgart, Germany.}
\date{\today}
\maketitle
\begin{abstract}
A method is presented to obtain {\it ab initio} upper and lower bounds to
surface energies of stress-driven discommensurate surface structures,
possibly non-periodic or exhibiting very large unit cells. The instability
of the stressed, commensurate
parent of the discommensurate structure sets an upper bound to its surface
energy; a lower bound is defined by the surface energy of an ideally
commensurate but laterally strained hypothetical surface system.
The surface energies of the phases of the Si(111):Ga and Ge(111):Ga systems
and the energies of the discommensurations are determined within $\pm 0.2$ eV.
\end{abstract}
%
%\pacs{PACS numbers: 68.35.Md, 68.35.Bs, 82.65.Dp}
\pacs{ }
]

  In the present paper we show how surface energies of discommensurate
surface reconstructions, which exhibit extraordinary large unit cells or
are not periodic at all, can be determined with an accuracy of a few tenths
of an eV using {\it ab initio} calculations of much simpler structures.
The method is applied to the Si(111):Ga and Ge(111):Ga systems, which
display several discommensurate reconstructions.

  Reconstructed surfaces can be classified according to their driving force,
which is most frequently chemical.\cite{chemdriv}
When the chemistry is satisfied, surfaces can still be unstable towards
rearrangements on a larger scale driven by other factors like
enthropy\cite{entdriv} or surface stress.\cite{vander,strdriv,vanderau}
Stress-driven discommensurate surface structures can be found for clean
surfaces, like Au(111),\cite{au} and adsorbate systems.  A rich
phenomenology has been found for metals adsorbed on (111) surfaces
of Si\cite{gasi} and Ge\cite{gage} for coverages in the monolayer range.

  An instability towards discommensuration arises when the parent structure
(the structure prior to discommensuration) is under a sufficient
compressive (tensile) surface stress.\cite{vander} A domain superstructure
appears at the surface. Within domains the surface partially
relaxes its stress by expanding (contracting) the surface lattice.
The structure within domains
resembles that of the parent structure whereas the domain walls,
which allow the adjustment to the subsurface periodicity,
represent defects. They are also considered as excitations if the
discommensurability occurs during a commensurate-incommensurate phase
transition.\cite{entdriv}

  The discommensurate structures of adsorbates on semiconductor surfaces
studied here display several characteristic differences with respect to
the better known physisorbed, discommensurate systems,
like noble gases physisorbed on graphite,\cite{stephens}
which are understood in terms of the
Frenkel-Kontorova model,\cite{FK} or with respect to the case of
Au(111), which is also understood in those terms and considerations of
continuum elasticity theory.\cite{strdriv,vanderau} The main differences are:
(i) The adsorbate chemisorbs mediated by covalent, i.e., directional bonds,
which represents a dramatic change of the energy scale. (ii) The
chemistry changes drastically at the domain walls. (iii) The
coexistence of two discommensurate phases at certain coverages suggests
a first order phase transition as function of coverage\cite{gage2}
[see Fig. 1~(a)]. These differences indicate the need for a
new qualitative understanding. As a first step, we analyze the
energetics of these complicated systems.

   First-principles studies of reconstructed surfaces are always
computationally demanding. However, the situation for discommensurate
structures is worse since the surface periodicity is lost (or the unit cell
is very large) and the geometry and bonding at domain walls
may not be well characterized.
Here we show how to obtain surface energies of
stress-driven discommensurate surface structures, using only information
obtained from commensurate structures of the system.
We demonstrate this for the Si(111):Ga and Ge(111):Ga systems.
Each system shows two stable discommensurate phases, $\gamma$ and
$\beta$, very close in saturation coverages but with quite different
structures [cf. Fig. 1~(a)]. Both systems are very similar, i.e.,
$\gamma_{\rm Si}$ is very
similar to $\gamma_{\rm Ge}$, and $\beta_{\rm Si}$ very similar to
$\beta_{\rm Ge}$~.\cite{gasi,gage} However, they differ substantially in
the fact that a commensurate $\sqrt 3 \times \sqrt 3 R 30^{\rm o}$ phase
($\sqrt 3$ hereafter) is stable for Si(111):Ga but not for Ge(111):Ga.
This rich scenario makes these systems very interesting on their own, but
also very illustrative of the usefulnes of our approach.

  The first-principles method used here is based on the local density
approximation,\cite{lda} and is described in detail elsewhere.\cite{gasi}
A repeated slab geometry, 12 layers thick, six layers of vacuum, approximates
the semi-infinite geometry of the surface.
The surface energy (per surface unit cell $\Omega$) is defined
as\cite{meadeprl}
\begin{equation}
\label{es}
E_s=(E_{slab}- N_{subs} E_{subs}^{bulk} - N_{ad} E_{ad}^{atom})/2 ,
\end{equation}
where $E_{slab}$ is the energy of the slab system per slab
unit cell, $E_{subs}^{bulk}$ is the energy per atom of
the substrate (Si or Ge) in its bulk form, $E_{ad}^{atom}$ is
the atomic energy of the adsorbate species (Ga),
and $N_{subs}$ and $N_{ad}$ are the number of atoms of substrate and adsorbate
species, respectively, per slab unit cell. The factor 1/2
reflects the presence of two surfaces per slab.

   A stress-driven discommensurate structure is stabilized (versus
its parent) by the energy which is released as the surface is allowed to relax
laterally within domains.  However, the energy gain is reduced
by the discommensuration energy ($E_{dis}$), which is always positive.
Two terms contribute to $E_{dis}$: (a) The energy required for creating
the domain walls, which represent defects from a chemical point of view,
since they disturb the otherwise preferred atomic arrangement at the
surface.\cite{chem}
(b) The energy which is necessary to allow for the local shears that appear
for a strained surface on an unstrained bulk.\cite{strdriv} While the
exact values of these energy contributions are not accessible by todays
{\it ab initio} methods, we can calculate upper and lower bounds of the surface
energy of such discommensurate systems. We start with the latter.

   When the whole parent system,
bulk included (the slab in this case), is strained laterally
by $\epsilon$ [Fig. 1~(b)],
the energy of the system increases, but the surface
energy $E_s^{par} (\epsilon)$ decreases towards a minimum [Fig. 1~(c)]
at a strain value very close to the one observed experimentally within
domains, $\epsilon_{dis}$.
In the $\beta$ phases half of the domains show a stacking fault at the
surface, the other half do not.\cite{gasi,gage} In this case
$E_s^{par} (\epsilon_{dis})$ is half of the sum of the surface energies of both
strained parent structures (with and without stacking fault). Energies
are obtained from first principles by relaxing the geometry up to the
fifth layer for each value of the strain.\cite{gasi,gage} Note that
$E_{subs}^{bulk}$ in Eq. (\ref{es}) is the energy of the substrate bulk under
the same strain. Since the energies shown in Fig. 1 (c) refer to the surface
unit cell of the strained system $\Omega^\prime$ they have to be corrected by
the ratio of areas of $\Omega$ and $\Omega^\prime$, giving
$E_s^{low} = E_s^{par} (\epsilon_{dis}) / (1 + \epsilon_{dis})^2$.

   The lower bound is more precisely adjusted by taking into account
the small change of adsorbate coverage at the domain
walls $\delta \theta_w$ (negative in the case of light walls like for
Si:Ga and Ge:Ga). This is done by adding the term
$E_s^{par} (\epsilon_{dis}) \delta \theta_w / \theta_{par}$, where
$\theta_{par}$ is the Ga coverage of the parent structure.
Since $1 / (1 + \epsilon_{dis})^2$ = $(\theta_{dis} - \delta\theta_w) /
\theta_{par}$, where $\theta_{dis}$ is the Ga coverage of the
discommensurate structure, we obtain the following lower bound:
\begin{equation}
\label{low}
E_s^{low} = E_s^{par} (\epsilon_{dis}) { \theta_{dis} \over \theta_{par}}.
\end{equation}
The energy of a discommensurate phase $\chi$ is then expressed as
\begin{equation}
\label{edisc}
E_s^{\chi}=E_s^{low,\chi}+E_{dis}^{\chi} .
\end{equation}

   For Si:Ga \cite{gasi} the coverages \cite{warncov} are
$\theta_\gamma=0.8$ monloayers (ML) and
$\theta_\beta=0.9$ ML, and $\theta_\gamma=0.7$ ML
and $\theta_\beta=0.8$ ML for Ge:Ga.\cite{gage} For the strain we use
$\epsilon_{dis} \approx 10 \%$,
similar to the experimentally observed value\cite{gasi,gage} and in
accordance with the minima in Fig. 1 (c).

   Upper bounds are obtained by an analysis of the surface energy
as a function of adsorbate coverage, using the fact that the
unstrained, commensurate parent structures are not observed experimentally.
The stability of the different phases
is shown by the {\it convex-hull} criterion:\cite{amitesh} in the graph
of energy versus coverage the lines joining the points representing
stable, saturated surface phases define a convex curve. This treatment
is strictly equivalent to the analysis based on the grand-canonical
free-energy as a function of adsorbate chemical potential,\cite{muchadi}
both at zero temperature. Required for this analysis and observed
experimentally are well defined saturation coverages for all the phases
involved.

   The only variable-coverage Ga phase that grows on the (111) surfaces of
Si and Ge is bulk Ga in the form of clusters. Assuming
bulk-like, large clusters, their internal energy per
Ga atom, $\mu_{\rm Ga}$, can be approximated by the corresponding value in
bulk Ga, $\mu_{\rm Ga}^{bulk}$.\cite{muchadi} This quantity has been
calculated for the
stable (orthorhombic) bulk Ga phase using the same technique and approximations
as for the surface calculations, giving $\mu_{\rm Ga}^{bulk}=-3.45$ eV/atom
($\mu_{\rm Ga}^{atom}=0$). In the plot of surface energy versus coverage,
the cluster phase appears as a straight line with slope equal to
$\mu_{\rm Ga}^{bulk}$. The line starts at the point associated with the
stable adsorbate phase of highest coverage, since the  surface energy as
defined in Eq. \ref{es} is  now the one corresponding to the adsorbate phase
plus the cluster energy which is equal to $\mu_{\rm Ga}^{bulk}
(\theta_{tot}-\theta_{ad})$. It should be noted that,
depending on the growth conditions,
Ga forms {\it small} clusters for coverages up to a few ML
which may increase the effective $\mu_{\rm Ga}^{clus}$ (a kinetic effect).

   To fullfil the convex-hull criterion,  the curve which joins the points
corresponding to saturation phases and ends in a line of slope
$\mu_{\rm Ga}^{bulk}$ has to be convex.
It is more convenient to use in the diagrams the reduced energy
\begin{equation}
\label{redes}
{\cal E}_s=E_s - \mu_{\rm Ga}^{bulk} \theta_{\rm Ga}.
\end{equation}
The stability curve remains convex, but the cluster line
has now zero slope (inset in Fig. 2).
The energies for the clean, reconstructed (111) surfaces
of Si\cite{brommer} and Ge\cite{selloni}, and $\sqrt 3$ phases for
Si\cite{meadeprl} and for Ge\cite{kuncproc} have been taken from the
literature.\cite{correct}

   In Fig. 2 convex-hull diagrams are shown for
Si:Ga and Ge:Ga. The bars indicate the energy range of stability for the
$\gamma$ and $\beta$ phases,
without taking into account the experimentally determined (in)stability of
the $\sqrt 3$ phases. The resulting surface energies are
$E_{s}^{\gamma}=-1.90 \pm 0.32$ eV/$\Omega$ and
$E_{s}^{\beta}=-2.25 \pm 0.32$ eV/$\Omega$ for Si:Ga, and
$E_{s}^{\gamma}=-1.71 \pm 0.23$ eV/$\Omega$ and
$E_{s}^{\beta}=-2.05 \pm 0.22$ eV/$\Omega$ for Ge:Ga.
%\cite{errors}.
The $\pm$ gives the margins of stability for these systems which are larger
than typical errors of full first-principles calculations.
However, the margins diminish with increasing domain size, since the strain is
smaller. This reduces the difference in energy between parent and
strained surfaces, the local shears, and the domain-wall energy per surface
atom (lower density of domain walls). Thus, in particular
for very large domains, where full first-principles calculations become
impractical, the applicability and accuracy of our method improve.
Furthermore, the {\it relative} energies of the discommensurate phases
are more precisely defined by the convex hull criterion: ${\cal E}_s^{\beta}$
has to be within [${\cal E}_s^{\gamma}-\delta$,${\cal E}_s^{\gamma}$], and
vice-versa, ${\cal E}_s^{\gamma}$ has to be
within [${\cal E}_s^{\beta}$,${\cal E}_s^{\beta}+\delta$], with $\delta$
$\leq 0.08$ eV/$\Omega$ in this case.

   We obtain important information about the discommensuration energy
$E_{dis}$ which is the energy increase of the discommensurate phase
compared to the lower bound (cf. Eq. \ref{edisc}). As indicated
in Fig. 2, $E_{dis}^{\gamma} > E_{dis}^{\beta}$ by at least 0.10
eV/$\Omega$ for Si and 0.14 for Ge [at most,
$E_{dis}^{\gamma} > E_{dis}^{\beta}$ by $(0.10+\delta) $
eV/$\Omega$ for Si, and by $(0.14+\delta)$ eV/$\Omega$ for Ge]. This means
energy differences of the order of 5 to 10 eV per domain. Taking into
account that the contribution of the local shears to these energies is
expected to be larger for $\beta$ than for $\gamma$ (larger domains), these
results indicate that we have chemically a better bonding
situation at the domain walls for
$\beta$ than for $\gamma$, both for Si and Ge. This agrees with the experimetal
findings of scanning tunneling microscopy measurements for Ge:Ga, showing
saturated bonds, i.e. little density of states in the band gap region at the
walls for $\beta$ but not for $\gamma$.~\cite{gage2} This
is actually the factor stabilizing the $\beta$ phase, a phase which is
{\it higher} in energy {\it within} domains due to the presence of the
stacking fault [cf. Fig. 1 (c)].

   The problem of the (in)stability of the $\sqrt 3$ phases is also addressed
in this analysis.\cite{correct} Figure 2 shows that the $\sqrt 3$ phase in
Si:Ga is stable, and that in Ge:Ga it is probably unstable (stable only if
the energies for $\gamma$ and $\beta$ would be at about their
upper limits).\cite{correct} Experimentally, the $\sqrt 3$ phase has been
shown to be stable for Si but not for Ge, though it can be prepared
for Ge at low temperature on cleaved (111) surfaces.\cite{sq3matthias}

   The surface energies of the discommensurate phases can be restricted to a
narrower range if we use the information about the (in)stability of the
$\sqrt 3$ phases as input\cite{correct} (insets in Figure 2).
The new surface energies are then
$E_{s}^{\gamma}=-2.05 \pm 0.18$ eV/$\Omega$ and
$E_{s}^{\beta}=-2.39 \pm 0.18$ eV/$\Omega$ for Si:Ga, and
$E_{s}^{\gamma}=-1.76 \pm 0.18$ eV/$\Omega$ and
$E_{s}^{\beta}=-2.10 \pm 0.18$ eV/$\Omega$ for Ge:Ga with the $\pm$ giving
again the margins.
%\cite{errors}.

   For Ge:Ga and approximately 1.5 ML Ga coverage, a third discommensurate
structure, $\beta_{H_3}$, has been characterized experimentally and observed
to be metastable.\cite{gage} It consists of Ga adatoms covering the $\beta$
phase at $H_3$ positions. The lower bound is calculated as for
the other structures and is found to be ${\cal E}_s^{low,\beta_{H_3}}=0.84$
eV/$\Omega$. It is just on the limit of stability (see Fig. 2). The surface
strain present in the underlying $\beta$ structure is an essential ingredient
for the appearance of $\beta_{H_3}$ versus other possible adlayer structures
(see Fig. 8 in Ref. \onlinecite{gage}). Its existence can be
understood in terms of an energy barrier toward the formation of
large Ga clusters (see above) and may furthermore suggest that the
energies for $\gamma$ and $\beta$ are closer to their upper than to their
lower bounds. As discussed in detail earlier,\cite{gage}
we believe, that this structure mediates an easy
transition from $\gamma$ to $\beta$.

   In summary, we have shown how to assess the
surface energy of discommensurate surface phases. Based on first-principles
calculations of much simpler, commensurate structures and on stability
considerations we obtain upper and lower bounds. We have demonstrated this
approach for Si:Ga and Ge:Ga.
The (in)stability of the $\sqrt 3$ phases has been also addressed
and values for the energies of the discommensurations
have been obtained. This information is particularly
important for understanding the phase diagram of these systems.

  We thank K. Kunc for communicating his results prior to publication.  E.~A.
is indebted to A.~Garc\'{\i}a, C.~Elsaesser, and S.~G.~Louie for the
permission to use their codes and for their help and support. E.~A.
acknowledges financial support from the Alexander von Humboldt Foundation
and from the DGICyT of Spain through grant PB92-0169.
%
% REFERENCES
%

\begin{figure}
\caption[Energy versus strain]{(a) Scanning tunneling microscopy image of
the $\gamma$ (left) and $\beta$ (right) phases coexisting on Ge(111):Ga
(taken from Ref. \onlinecite{gage2}). (b) Commensurate, strained
commensurate, and discommensurate structures in side view.
(c) Surface energy per {\em strained} surface unit cell
($\Omega^\prime$) versus lateral strain for Ge(111):Ga-(1$\times$1) with Ga
substituting the surface layer, with (full circles) and without (empty
circles) stacking fault at the surface. D.W. means domain wall.}
\label{estrain}
\end{figure}

\begin{figure}
\caption[Stability diagram 1]{Stability diagrams for Si:Ga and Ge:Ga.
Diamonds indicate unstrained commensurate structures, filled
circles indicate lower
bounds for $\gamma$ and $\beta$ structures (and for $\beta_{H_3}$ in Ge:Ga),
bars indicate their ranges of stability, and squares indicate
$\sqrt 3$ structures. The insets show the diagrams in terms of the reduced
energy (cf. Eq. 4). Arrows indicate ranges of stability if
exploiting the knowledge about the (in)stability of the $\sqrt 3$  phases.}
\label{diag1}
\end{figure}


\begin{references}
\bibitem{chem} If this term is negative the chemistry at
the domain walls would be preferable to the one within domains and a chemical
reconstruction of a smaller scale would appear.
\bibitem{warncov} There is some uncertainty in the experimental
determination of the coverages. However, the accuracy of the
values is sufficient for the purposes of this paper.
\end{references}
\end{document}